\newcommand{\CLASSINPUTtoptextmargin}{0.75in}
\newcommand{\CLASSINPUTbottomtextmargin}{1.01in}
\documentclass[letterpaper, conference]{IEEEtran}
\IEEEoverridecommandlockouts
\usepackage{cite}
\usepackage{amsmath,amssymb,amsfonts}
\usepackage{algorithmic}
\usepackage{graphicx}
\usepackage{textcomp}
\usepackage{xcolor} \usepackage{fancyhdr}
\def\BibTeX{{\rm B\kern-.05em{\sc i\kern-.025em b}\kern-.08em
    T\kern-.1667em\lower.7ex\hbox{E}\kern-.125emX}}

 \fancypagestyle{firstpage}{%
 \fancyhf{}%
\fancyhead[L]{IEEE Military Communications Conference\\
 \textit{October 28 - November 01, 2024, Washington, DC, USA}}

  \fancyfoot[L]{%
\normalsize{979-8-3503-8359-1/24/\$31.00 ~\copyright2024 IEEE}
}%
 }

\begin{document}

\title{
    {Trust or Bust: Ensuring Trustworthiness in Autonomous Weapon Systems}
}

\author{\IEEEauthorblockN{1\textsuperscript{st} Kasper Cools}
\IEEEauthorblockA{
\textit{Belgian Royal Military Academy,} \\
\textit{
    Vrije Universiteit Brussel} \\
    kasper.cools@mil.be}
\and
\IEEEauthorblockN{2\textsuperscript{nd} Clara Maathuis}
\IEEEauthorblockA{
\textit{
    Open University of the Netherlands} \\
    clara.maathuis@ou.nl}
}

\maketitle
\thispagestyle{firstpage}

\begin{abstract}
The integration of Autonomous Weapon Systems
(AWS) into military operations presents both significant opportu-
nities and challenges. This paper explores the multifaceted nature
of trust in AWS, emphasising the necessity of establishing reliable
and transparent systems to mitigate risks associated with bias,
operational failures, and accountability. Despite advancements in
Artificial Intelligence (AI), the trustworthiness of these systems,
especially in high-stakes military applications, remains a critical
issue. Through a systematic literature review, this research
identifies gaps in the understanding of trust dynamics during
the development and deployment phases of AWS. It advocates
for a collaborative approach that includes technologists, ethicists,
and military strategists to address these ongoing challenges. The
findings underscore the importance of Human-Machine teaming
and enhancing system intelligibility to ensure accountability and
adherence to law. Ultimately, this paper aims to contribute to
the ongoing discourse on the ethical implications of AWS and
the imperative for trustworthy AI in defense contexts.\end{abstract}


\section{Introduction}
Building AI systems in a trustworthy manner is fundamental to their successful integration into society and various industries, including the defence sector. Recent studies indicate that over $70\%$ of individuals express neutral or no confidence in AI systems, underscoring the critical need for establishing trust. This is particularly relevant in the context of Autonomous Weapon Systems (AWS), where the stakes are high and the margin for error is low~\cite{LethalAWS-US, TAI-XAI, TAI-Principles}. While various research studies are dedicated to understanding the challenges and opportunities associated with AWS, there remains a noticeable gap in the literature concerning the trustworthiness of these systems. This research aims to bridge this knowledge gap. \\ 
Trustworthy AI systems, including AWS, are essential for mitigating potential risks, such as deception, privacy invasion, and perpetuation of biases, which could lead to catastrophic outcome. To achieve this, a multifaceted approach is required, encompassing several key aspects. First, developers must focus on promoting fairness and reducing bias by implementing measures to detect and correct biases in training data. Second, enhancing security through robust measures to protect against attacks is crucial for safeguarding personal data and system integrity. In the case of AWS, this also means protecting against adversarial attacks that could compromise the system's functionality. Third, improving explainability by enabling AI systems to provide understandable explanations for their decisions enhances transparency and user trust. Lastly, ensuring compliance with regulatory requirements is vital for the legal and safe use of AI systems. For AWS, adherence to international law and norms is paramount~\cite{LethalAWS-UN, AWS_DESIGN, TRUSTWORTHY_AI, TAI-XAI, TAI-Principles}. By addressing these factors, developers can create AI systems, including AWS, that are not only technologically advanced, but also socially responsible and widely accepted, ultimately building public confidence and promoting the ethical, safe, and effective use of AI systems ~\cite{TAI-XAI, TAI-Principles, LethalAWS-US}.

This paper proceeds as follows: first, the related studies to this research are discussed in Section~\ref{section:related-work}. Subsequently, the research methodology is detailed in Section~\ref{section:research-methodology}. In Section~\ref{section:challenges}, the challenges associated with AWS are discussed, followed by an exploration of the opportunities AWS presented in Section~\ref{section:opportunities}. Finally, we conclude our findings in Section~\ref{section:conclusion}.

% + outline

\section{Related Work\label{section:related-work}}
%C: related studies for this article.
In the context of AI, autonomy refers to a system's ability to make decisions and act independently without human intervention. This capability extends beyond simples automation, as autonomous AI systems can adapt their actions based on learning from their environment and situational awareness. For AWS, autonomy is characterised by the ability to select and engage targets independently, based on predefined criteria and environmental interpretation. AWS operate in complex environments requiring rapid decision-making. They use self-learning capabilities to adapt over time, navigating unpredictable scenarios without constant reprogramming. Nevertheless, the definition of AWS also considers the degree of human control, which varies among different interpretations. Some emphasise maintaining human oversight in critical decision-making processes, while others focus on the system's independent operation. Balance between autonomy and human control is a central point of debate in the ethical and legal considerations surrounding AWS~\cite{AWS-Definitions}. 

To address this,~\cite{AWS-Systemic} proposes adopting a systemic view when building and using AWS. This involves a comprehensive approach that recognises AWS as complex entities within larger systems supported by three fundamental pillars: system, autonomy, and values. The system pillar views AWS as self-regulating components interacting with humans, laws and ethics. The autonomy pillar highlights AWS’s independent operation based on programming and data. The values pillar addresses the legal and moral considerations governing AWS use, ensuring compliance with International Humanitarian Law (IHL) and ethical standards. This holistic approach acknowledges the inherent complexity of AWS and the challenges in their regulation. By adopting a systemic vision, stakeholders can better understand the technological, operational, legal, and ethical dimensions of AWS, leading to more effective development, deployment, and regulation strategies. Key aspects of regulation include the need for updated and comprehensive guidelines that go beyond existing frameworks. The Advisory Council on International Affairs and the Advisory Committee on Issues of Public International Law advocate for more concrete regulations that cover the development, procurement, and use of AWS, emphasising the importance of clear criteria and responsibilities for all stakeholders involved. Additionally, international cooperation involving intensified consultations among governments, the private sector, and civil society is necessary to ensure that AWS development remains within ethical and legal boundaries. The involvement of a broad range of stakeholders is crucial to creating effective, widely accepted, regulatory frameworks that address the potential for abuse and ensure compliance with IHL~\cite{AWS-Regulation, AWS-Law, AWS-UN}. \\
In particular, Lethal Autonomous Weapon Systems (LAWS) pose ethical, legal, and operational challenges. Ethically, the primary concern is the moral acceptability of machines making life-and-death decisions, which challenges the principles of distinction and proportionality required by the law of war. Accountability is another major issue, as determining responsibility for unlawful actions or malfunctions becomes complex when autonomous systems are involved. Operational risks, including the potential for malfunctions or exploitation by adversaries, further complicate the deployment of LAWS, necessitating robust safeguards to minimise failures. Compliance with international law is also critical, as there are significant doubts about whether AWS can adhere to legal standards in complex combat scenarios. Finally, the absence of international consensus on LAWS’ definition and regulation hinders control measures, given diverse views on their usage and the possibility of a ban or strict rules. These issues reveal the need for careful consideration and comprehensive regulation to address the ethical, legal, and operational challenges posed by LAWS~\cite{LethalAWS-UN, LethalAWS-US}. \\
A perspective that allows tackling some of the existing issues is forming and engaging in human-autonomy teaming (HAT). In the context of military operations, this refers to the collaboration between human operators and autonomous systems to achieve common goals, leveraging the strengths of both. This concept is crucial in enhancing operational capabilities, as autonomous systems can process large volumes of data rapidly and operate in hazardous environments, providing superior situational awareness and decision-making support. Trust between humans and autonomous systems is foundational to effective HAT, ensuring that operators are comfortable relying on AI for proactive decisions in dynamic battlefield conditions. Cultural and psychological factors, such as automation bias, influence trust and must be understood to maximise the benefits of AI integration. Additionally, HAT must operate within strategic, political, and legal frameworks, adhering to international laws of armed conflict. The continuous advancement of AI offers new opportunities, making the successful implementation of HAT essential for maintaining operational effectiveness and strategic advantage~\cite{HAT-MO, HAT-MO-Prop}. \\
It is then important, as~\cite{AWS-ChallengesOpportunities, AWS-ProsCons} argue, to understand both the challenges and opportunities as well as the (dis)advantages of building and using AWS. On the positive side, AWS drive innovation, enhance defence capabilities, and offer strategic advantages by operating in high-risk environments with increased precision and speed. They have the potential to reduce human casualties and increase operational efficiency through rapid data processing and independent task execution. However, AWS also face significant challenges, including technological and financial barriers, cyber security risks, and critical moral and legal concerns. The ethical implications of entrusting life-and-death decisions to machines, potential violations of international laws of war, and issues of accountability are paramount. Additionally, the rapid pace of technological advancement complicates effective regulation and definition of AWS, while potentially exacerbating inequalities between nations due to dependencies on foreign technology. Balancing these pros and cons is essential for responsible development and deployment, requiring careful consideration of ethical, legal, and operational implications to ensure alignment with international laws and moral standards. Hence, in current and future applications, proper governance of AWS is crucial to address responsibly and safely the complex interplay of ethical, legal, security, and technological challenges they present, both now and in future applications. Effective governance frameworks are crucial to ensure AWS operate ethically and legally, especially in terms of accountability and compliance with IHL. As AWS has the potential to transform military strategies, governance is needed to manage the risks associated with their deployment, including escalation of conflicts and the proliferation of these technologies to malicious actors. Moreover, governance of AWS must account for the political and economic implications of their development, balancing national security interests with the potential for destabilisation in international relations. Establishing clear guidelines for the production, deployment, and use of AWS is vital to maintain human oversight and accountability, ensuring decisions regarding lethal force remain under human control~\cite{AWS-Gov, AWS-FutureGov}.\\
Nevertheless, as this extensive review shows, while various academic and practitioner studies are dedicated to capturing various dimensions characterising the processes of building and using AWS in military operations, a systematic approach to understand and be able to further materialise corresponding challenges and opportunities in the military domain is lacking. This represents the knowledge gap that this article aims to tackle in this study.

\section{Research Methodology\label{section:research-methodology}}
%C: presentation of the research methodology taken in this article.
This research aims to analyse the challenges and opportunities encountered when building and using AWS. On this behalf, the following research questions are formulated:
\begin{itemize}
    \item Research Question 1: What are the challenges associated with the development and utilisation of Autonomous Weapon Systems?
    \item Research Question 2: What opportunities are associated with the development and utilisation of Autonomous Weapon Systems?
\end{itemize}

To answer these research questions, a systematic literature review approach is considered following the well-known PRISMA methodology~\cite{PRISMA}. This implies identifying relevant research studies, selecting the relevant literature based on inclusion and exclusion criteria, and conducting an in-depth analysis and presentation of the findings gathered. Specifically, the search string included keywords like Automated Weapon Systems, trustworthy, safety, responsibility, military, Artificial Intelligence, Machine Learning, and Deep Learning. The search strings were used to query the IEEE Xplore, ACM Digital Library, Taylor Francis, Sage Journals, and Wiley scientific databases. As inclusion criteria, only journal and conference studies, in English, and between 01.01.2013 and 31.12.2023, are considered. This resulted in a total of $10.354$ studies after which their abstracts were reviewed based on exclusion criteria such as duplicates and studies that are not dedicated to the military domain. This resulted in a refined list of $629$ articles. A thorough content analysis was then performed to identify the ones that satisfied our criteria, articles shorter than 3 pages were excluded. 
From this process, $45$ articles are considered for inclusion in the final step, which represent the ones that are in-depth analysed in this research. Accordingly, the list with the aim of these studies together with their type (i.e., review, survey, position, or application) is presented in Table ~\ref{table:articles-reviewed}.
\begin{table*}
    \caption{Articles Reviewed\label{table:articles-reviewed}}
    \begin{center}
        \tabcolsep=0.11cm
        \begin{tabular}{|c|c|c|}
            \cline{1-3}
            \textbf{\#} & \textbf{\textit{Aim}}                                                                                                            & \textbf{\textit{Type}} \\
            \hline
            \cite{P1}   & To analyse types of trust required for assuring ethical use of AWS.                                                              & Review                 \\
            \cite{P2}   & To understand the dynamics between LAWS and Human-Machine Teaming.                                                               & Position               \\
            \cite{P3}   & To understand the levels of trust in Human-Autonomy Teaming in military operations.                                              & Position               \\
            \cite{P4}   & To analyse the mechanisms of trust evaluation in the context of Human-Machine Interaction.                                       & Review                 \\
            \cite{P5}   & To understand design factors of AWS in relation to human control.                                                                & Position               \\
            \cite{P6}   & To understand the notion of human control in the context of using autonomous systems in the military and transportation domains. & Review                 \\
            \cite{P7}   & To investigate the effect of promises on the trust repair process in the context of autonomous systems.                          & Survey                 \\
            \cite{P8}   & To analyse key psychological insights of Human-Machine Interactions in AI-enabled warfare                                        & Position               \\
            \cite{P9}   & To assess the potential development and deployment of LAWS through security studies lenses.                                      & Review                 \\
            \cite{P10}  & To understand the meaning of moral injury when using AWS.                                                                        & Position               \\
            \cite{P11}  & To propose a reputation management framework for UAV to check the realness of event messages.                                    & Application            \\
            \cite{P13}  & To propose a blockchain AI-based solution for identification and tracking illegal UVAF in an IoMT system.                        & Application            \\
            \cite{P14}  & To propose an ethical assessment framework for emerging AI and robotics technologies.                                            & Application            \\
            \cite{P15}  & To review existing research studies focusing on situational awareness for autonomous transport.                                  & Review                 \\
            \cite{P16}  & To analyse common high-level security requirements of autonomous systems in missions.                                            & Position               \\
            \cite{P17}  & To propose a trust loss effects analysis methodology for systematic risk assessment of autonomous systems                        & Application            \\
            \cite{P18}  & To analyse existing trends in relation to the development and use of AWS.                                                        & Review                 \\
            \cite{P19}  & To propose a sound concept development trajectory model for UAV.                                                                 & Application            \\
            \cite{P20}  & To review principles of designing multi-team collaboration systems.                                                              & Review                 \\
            \cite{P21}  & To present autonomous intelligent solutions for fighting autonomous intelligent malware as a potential future of cyber defence.  & Survey                 \\
            \cite{P22}  & To propose a regulatory framework for autonomous systems.                                                                        & Application            \\
            \cite{P23}  & To reflect on the definition of AWS, the underlying technologies, and integration into the future force.                         & Review                 \\
            \cite{P24}  & To analyse the American public attitudes towards AWS.                                                                            & Survey                 \\
            \cite{P25}  & To discuss the underlying ethical dimensions of AI-enabled military drones.                                                      & Position               \\
            \cite{P26}  & To reflect on the underlying moral dimensions and implications of Autonomous Systems.                                            & Position               \\
            \cite{P27}  & To understand the role of AI in kinetic targeting and the compliance with IHL.                                                   & Position               \\
            \cite{P28}  & To reflect on the role and implications that AI technologies and systems have in the context of nuclear stability.               & Position               \\
            \cite{P29}  & To reflect on the non-descriptivism and descriptivism judgements of AWS.                                                         & Position               \\
            \cite{P30}  & To reflect on the applicability of deterrence theories in the age of AI and autonomy in relation to nuclear deterrence.          & Position               \\
            \cite{P31}  & To analyse regulation perspectives of military AI systems through EU lenses.                                                     & Position               \\
            \cite{P32}  & To review security issues applicable to IoD (Internet of Drones).                                                                & Review                 \\
            \cite{P33}  & To reflect on human control and the use of force in the context of AWS.                                                          & Position               \\
            \cite{P34}  & To reflect on human control and responsibility when using AWS.                                                                   & Position               \\
            \cite{P35}  & To assess regulatory challenges and opportunities that killer robots have.                                                       & Position               \\
            \cite{P36}  & To analyse the types of attacks and corresponding defence mechanisms in the context of AWS.                                      & Position               \\
            \cite{P37}  & To analyse the advantages and disadvantages of the deployment of AWS.                                                            & Position               \\
            \cite{P38}  & To reflect on the components and corresponding sub-components of UVS aircrafts.                                                  & Review                 \\
            \cite{P39}  & To reflect on the ethical and moral use of LAWS.                                                                                 & Position               \\
            \cite{P40}  & To reflect on the legal implications of UAVs in relation to the potential revisions of LOAC related to autonomous systems.       & Position               \\
            \cite{P41}  & To analyse US's decision-making and moral justifications on the use of AWS.                                                      & Position               \\
            \cite{P42}  & To analyse the meaning of responsibility of autonomous robots.                                                                   & Position               \\
            \cite{P43}  & To reflect on the legal implications and consequences of AWS in the theatre of war.                                              & Position               \\
            \cite{P44}  & To analyse media discourses related to AWS.                                                                                      & Position               \\
            \cite{P45}  & To reflect the meaning and implications of the applicability of IHL in respect to the use of AWS.                                & Position               \\
            \cite{P46}  & To analyse the role and necessity of involving humans in missions where AWS are used.                                             & Position               \\
            \cite{P47}  & To build a model that assesses the reliability of the military UAV in the presence of external attacks.                            & Application            \\
            \hline
        \end{tabular}
        \label{tab1}
    \end{center}
\end{table*}

\section{Challenges of AWS\label{section:challenges}}
This section highlights key challenges in AWS development and deployment, examining potential hurdles in modern warfare and defence strategy, categorised into five clusters:

Challenge 1 - System reliability:
    Even with a strong reliability framework, the operational reliability of AWS cannot be fully assured due to unpredictable interactions between subsystems and the unexpected situations they might face during military operations~\cite{P47}. It is important to note that most software, including those used in AWS, can be insecure, often due to the exclusion of security requirements from the specifications. This insecurity, coupled with the decreasing cost of hardware for brute-force attacks, complicates the issue of key management. Therefore, it becomes essential to continuously validate and enhance their reliability, with an emphasis on security aspects, and address these challenges to enhance the trustworthiness of AWS~\cite{P6, P19, P27, P20, P21}. In the context of military operations, AWS can leverage various types of wireless sensors distributed before battle. These sensors, connected to autonomous software agents in a multi-agent system, can provide an on-field tactical advantage. However, ensuring the integrity of communications between these sensors is crucial. Public key infrastructure could be a solution, but it brings its own challenges, such as secure storage for the private key and secure over-the-air transmission of keys~\cite{P11}. Regarding AWS, trust is a binary decision that hinges on whether one views them as mere tools, where reliability and predictability are sufficient, or as moral agents with higher trust thresholds. Addressing the significant technological challenges in testing, validating, and verification of these systems for predictability and reliability is crucial \cite{P27, P1, P38}. AI capabilities have seen significant improvements, especially in object recognition, but still face challenges in complex environments. Deep learning models, for instance, excel at learning and adapting independently, but may struggle with tasks that deviate from their training, and they lack human-like contextual understanding. Furthermore, AI systems can be `brittle', struggling with tasks that slightly differ from their design specifications. These models are often perceived as `black boxes`, making it difficult to comprehend their decision-making processes, which is problematic in situations requiring decision transparency~\cite{P34, P46, P31}. When deploying AWS, it is essential to establish well-defined accountability mechanisms to prevent the emergence of a system characterised by organised irresponsibility. Moreover, attempts to translate accountability rules such as Rules of Engagement and IHL into software have mostly been unsuccessful due to the complexity involved. As of today, there is no practical method available to mimic the qualitative assessments involved in analysing the proportionality of a military action.~\cite{P37, P18, P22}

Challenge 2 - Training and Adaptability:
    Another key challenge is managing the balance of trust between humans and AWS. It is crucial to ensure that human actors neither over- nor under-trust these systems. Despite their exceptional adaptability, humans often misjudge a machine's capabilities, leading to `automation bias'. This can manifest in two ways: humans might over-trust the system, accepting its decisions even when they are unjust, or they might under-trust it, constantly monitoring or overriding the system's tasks ~\cite{P1, P2, P14}.
    
    In addition, while AWS have the potential to reduce stress and cognitive overload, their effectiveness is constrained by their ability to gather and convey accurate information relevant to their tasks. This is primarily due to the current state of AI capabilities. At present, AI is not as proficient as humans at recognising objects, especially in complex environments. Furthermore, despite potential advancements in image recognition, AI exhibits a fundamental `brittleness'. They are trained to perform specific, narrowly defined tasks and face difficulties when adapting to tasks that are even slightly different ~\cite{P46}.

Challenge 3 - Strategic vulnerability:
    One of the key issues is the concept of first-strike instability, which emerges as a strategic vulnerability in the development and deployment of AWS. This instability arises from the inherent characteristics of AWS, such as their speed, precision, and the ability to limit human risk. These attributes potentially incentivize a preemptive strike, offering tactical advantages like rapid and accurate attacks at a lower cost, without endangering personnel. This scenario, where one party is motivated to strike first due to these advantages, constitutes first-strike instability, even as AI continues to evolve towards reduced brittleness and increased transparency ~\cite{P46}. This issue of first-strike instability is further complicated by the secretive nature of AWS, which can lead to increased uncertainty among allied nations. Specifically, nations might be reluctant to share details about AWS capabilities, which could potentially put a strain on relationships ~\cite{P46}.
    On a practical level, there are doubts about the ability of AI-enabled systems to accurately differentiate between legitimate and illegitimate targets. Moreover, the potential consequences of any misidentification are immense, further amplifying the complexity of the issue at hand~\cite{P23, P8}. The first-strike instability of AWS presents several challenges. Decision-making algorithms often lack transparency, leaving policymakers in the dark - a problem known as the `black box' challenge. Additionally, AWS faces data-related issues such as data acquisition, data skew, misuse, and privacy concerns. At the same time, the potential of AI to perpetuate existing discriminatory structures and biases is a significant concern that needs to be addressed. Lastly, the role of humans in these autonomous systems varies across contexts; in military systems, human oversight may be legally or ethically required, whereas, in transportation, human involvement might be less necessary - especially if human error is a major contributor to operational incidents~\cite{P22}. In the context of strategic vulnerability, accountability and responsibility are major challenges. It is essential to establish clear rules and mechanisms when using AWS to prevent irresponsible behaviour. However, attempts to translate laws into software have proven to be challenging, and no practical strategy exists to evaluate proportionality in complex military scenarios~\cite{P37}.
    
Challenge 4 - Human-Machine teaming:
    Establishing trust and understanding is a key challenge in Human-Machine teaming. AWS are dynamic systems that evolve and adapt, making them unpredictable, and requiring humans to have a deep understanding of how they work in order to be able to trust them fully.~\cite{P1, P2, P3}. The balance between operational autonomy and control also presents a challenge. System designers must strike a balance between giving robots enough autonomy to be useful, while still maintaining appropriate human control. Finding the right balance is key: humans must have an appropriate level of trust in autonomous systems, but not too much~\cite{P3, P41}.
    Subsequently, there is also training and adaptability. The degree to which a learning AWS truly adapts to its environment in operationally effective ways is inversely related to the degree to which human team members can recognise and use its data effectively. As AWS improves, its ability to adapt may decrease its usefulness when working with humans, because humans struggle to understand and utilise its enhanced capabilities~\cite{P1}.
    Lastly, communication and coordination are crucial for Human-Machine teaming. Policymakers perceive autonomy not as a fixed state of control, but as a layered process of collaboration between humans and computers. The Defense Science Board's perspective on autonomy, on the other hand, is unique in that it defines autonomy as the deliberate assignment of cognitive tasks and responsibilities between humans and computers to achieve certain capabilities. This operational-dynamic approach acknowledges that an AWS may have varying degrees of control over different mission aspects at any given time~\cite{P23}.

Challenge 5 - Capturing ethical, legal, and social dimensions:
    The push for incorporating ethical principles in AWS highlights the unresolved task of choosing the most fitting ethical standard for a situation. Various ethical theories have been used to tackle moral dilemmas. However, there is currently no consensus among ethical experts regarding the choice of ethical theory to apply~\cite{P44, P29, P8}. Establishing robust accountability mechanisms is essential when deploying AWS to prevent systemic irresponsibility. The risk of AI amplifying discriminatory structures is a significant concern, necessitating proactive measures to counteract potential bias~\cite{P37, P22}. Furthermore, the societal implications and public perception of AWS also present a significant challenge. Central issues include the AI potential to reinforce existing discriminatory structures and the democratic accountability of autonomous systems, particularly when these systems make decisions affecting public rights or the allocation of public benefits~\cite{P37, P27}. Subsequently, transparency and explainability are also significant challenges. A primary concern is the transparency of autonomous systems in terms of decision-making, often referred to as the `black box' challenge. Unlike the human brain, which continuously learns and builds upon previously acquired knowledge, algorithms are typically trained once on a dataset. This approach restricts their ability to learn new information without retraining, posing challenges in dynamic and complex environments where continuous learning is vital~\cite{P24, P14}. Finally, using autonomous systems in military contexts brings ethical dilemmas related to human rights and warfare principles into focus. The ability of humans to challenge the codified ethics of these machines could be reduced, leading to a moral vacuum where existing laws, ethical guidelines, and societal norms may be insufficient. On a more pragmatic note, there are doubts about whether AI-driven functionalities can accurately distinguish between lawful and unlawful targets ~\cite{P24, P26}.

    \section{Opportunities of AWS\label{section:opportunities}}
    Moving forward, this section explores the opportunities that the development and deployment of AWS can bring to modern warfare and defence strategies.
    The benefits and advantages of AWS are organised into five key areas:

Opportunity 1 - Capacity:.
\begin{itemize}
    \item Improved capabilities: AWS significantly advances military technology, offering improved offensive and defensive capabilities while addressing key challenges in modern warfare. These systems can navigate complex, dynamic environments, essential for modern tactical combat operations~\cite{P36}. One of the biggest advantages of AWS is their potential to reduce casualties through enhanced targeting precision. Advanced image processing and object detection capabilities, often surpassing human abilities, enable AWS to more accurately distinguish between legitimate military targets and non-combatants~\cite{P39, P28, P30}. This technology facilitates positive target identification through facial recognition and complex analysis of objects and behaviours, allowing for the detection of potential threats such as improvised explosive devices or suicide bombers~\cite{P27}. These advancements in autonomous systems suggest a future of warfare that is potentially more precise, ethical, and less costly in terms of human life.
    \item Enhanced decision-making: AWS provide enhanced decision-making capabilities and improved operational efficiency. These systems can process and synthesise vast amounts of data from multiple sources quicker than human operators, enabling faster and more accurate responses to threats~\cite{P27}. This speed advantage is essential in high-stake environments where quick reactions can determine the outcome of engagements~\cite{P6}. Furthermore, AWS can significantly accelerate the Observe-Orient-Decide-Act decision-making loop, allowing for more efficient execution of missions~\cite{P41, P42}. By operating at high speeds, AWS can outperform human commanders in coordinating complex military manoeuvres~\cite{P47}. Furthermore, these systems can maintain a higher level of accuracy and effectiveness over extended periods, as they do not suffer from fatigue or cognitive limitations~\cite{P7}. The autonomous operation of AWS also reduces the cognitive load on human operators, allowing them to focus on higher-level tasks and potentially decreasing the overall size of the military workforce~\cite{P7, P9, P15}. While AWS introduce new challenges, they are designed to make previously difficult tasks more manageable, ultimately balancing the cognitive demands placed on human operators~\cite{P46}.
    \item Safe and efficient resource allocation: Deploying AWS offers numerous advantages in resource allocation, planning, and operational safety. AWS can lead to significant cost savings over time, as machines do not require the logistical support needed for human soldiers, such as food, medical care, and leave, and can undertake high-risk missions without the associated human costs~\cite{P23, P24}. Decentralising command and control (C2) through AI enhances the operational efficiency of operations, allowing unmanned entities to execute more effective defensive strategies~\cite{P19}. Networked AWS platforms enable superior planning and coordination, as they can synchronise fire and manoeuvre more effectively than humans~\cite{P3}. The ability to operate with minimal human oversight allows AWS to function in complex networks, executing coordinated attacks or serving as mobile, self-healing minefields~\cite{P46}. Lightweight encryption and authentication protocols have the potential to protect sensitive data and maintain communication integrity between Unmanned Aerial Vehicles (UAVs) and ground control systems~\cite{P11, P15}. Moreover, remote operation capabilities, such as using UAVs for reconnaissance and armed attacks, allow operators to conduct missions from a safe distance, significantly reducing the risk to soldiers and enhancing overall safety~\cite{P38}.
    \item Enhanced ISR (Intelligence, Surveillance, and Reconnaissance) missions: AWS, and in particular UAVs, enable superior monitoring and intelligence gathering without endangering human lives~\cite{P11}. The effectiveness of UAVs in military surveillance and espionage has been well-established for over two decades, providing real-time data to military personnel~\cite{P15}. AWS employ sensor-based intelligence gathering, which offers several advantages, including easier deployment, camouflage capabilities, and enhanced reliability through redundancy~\cite{P36}. The compact size of these sensors allows for widespread deployment and concealment in various environments, providing a strategic edge in pre-engagement reconnaissance missions~\cite{P36}. Furthermore, autonomous systems can detect enemies, conduct surveillance, and potentially carry weapons for situations requiring lethal force, significantly enhancing military operational capabilities~\cite{P41, P42, P44}.
\end{itemize}

Opportunity 2 - Training and Awareness: AWS are being driven by the involvement of technical and corporate experts in their design, development, deployment and governance, ensuring the incorporation of cutting-edge technologies and providing a competitive edge in military operations~\cite{P31}. While fully autonomous weapon systems are not yet publicly deployed, military personnel are being prepared for their eventual implementation through cost-effective simulated training exercises, workshops, and tabletop waggles~\cite{P1, P5}. These preparatory measures are crucial for building familiarity and encouraging experimentation with AWS before operational deployment, as exemplified by the Australian Defence Force's current practices~\cite{P5}. Collaborative research between basic and applied scientists is enhancing the understanding of human-oriented aspects in autonomous systems, leading to improved AWS design and implementation~\cite{P14, P15}. Furthermore, insights gained from studying software contributions to military aviation and Remotely Piloted Aircraft Systems accidents are being applied to enhance the safety and effectiveness of AWS, informing the development of more stringent safety testing, certification, and regulation processes~\cite{P46, P48}.

Opportunity 3 - Strategic and Military Advantage: The development and integration of AWS into military operations present significant strategic advantages for technologically advanced nations, particularly in enhancing precision, speed, and efficiency~\cite{P1, P2}. This focus on machine autonomy is driven by the potential for improved operational effectiveness through Human-Machine teaming~\cite{P2}. AWS integration into C2 systems offers the prospect of enhanced decision-making processes, providing commanders with data-driven insights and a broader range of tactical options~\cite{P22, P25}. Public perception of AWS appears to be malleable, with increased support when presented with information highlighting their potential to save soldiers' lives and reduce errors~\cite{P22}.AWS enhance combat efficiency and mitigate risk by operating in high-threat environments, boosting capabilities while reducing risk to human life~\cite{P22}. These systems act as force multipliers, performing tasks that would otherwise require multiple human soldiers, thereby improving operational efficiency~\cite{P42, P45}. European weapon manufacturers have contributed to cumulative improvements in AWS through design choices and practices, shaping implicit rules for appropriate human control in specific use-of-force situations~\cite{P31, P32, P33}.

Opportunity 4 - Human-Machine teaming: The integration of AWS emphasises the critical importance of Human-Machine teaming, where trust between humans and machines is essential for operational success. Effective collaboration between humans and autonomous systems is essential to ensure reliability and performance in operational settings~\cite{P2}. As military strategies evolve, the boundaries between semi-autonomous devices and LAWS highlight the increasing reliance on this synergy, suggesting that future operations will depend heavily on effective Human-Machine interactions~\cite{P2}. The modernisation plans of various forces, such as the Italian army's focus on autonomy and the UK's investment in drone swarms, reflect a broader trend towards enhanced autonomy, potentially leading to more effective and efficient defence strategies~\cite{P18}. Furthermore, as AWS capabilities expand, the potential for improved decision-making and strategic flexibility becomes evident, enabling commanders to access a wider range of tactical options and insights. 

Opportunity 5 - Capturing ethical, legal, and social dimensions: The ethical, social, and legal considerations surrounding AWS are multifaceted and critical to their deployment. One of the primary ethical concerns is the potential to alleviate moral injury among soldiers by assuming the responsibility for lethal decisions, thereby reducing the psychological burden and guilt associated with combat actions~\cite{P9, P10}. This concept of `moral off-loading' raises complex questions regarding the delegation of moral responsibility, as it shifts the ethical implications of killing from the soldier to the machine~\cite{P9, P10}. To ensure the responsible use of AWS, the development and deployment of these systems must be guided by evolving standards and regulations. International organisations such as ISO, IEC, and IEEE are actively working to create technical standards that address critical issues like algorithmic bias and trustworthiness, which are essential for the acceptance of AWS in military contexts~\cite{P9}. Establishing a professional military code of ethics is also vital, as it addresses fundamental questions about design values and legal responsibilities, particularly in high-stakes environments~\cite{P13, P14}. Furthermore, the introduction of AWS necessitates a re-evaluation of existing legal frameworks to accommodate the unique challenges posed by autonomous systems. This re-evaluation presents an opportunity for innovation in military law and ethics, as seen in initiatives like the U.S. Department of Defence's Defence Innovation Initiative~\cite{P29}. Their potential to influence international norms and regulations is significant; they may alter how core provisions of international law are interpreted or operate alongside existing legal standards~\cite{P31, P33, P34, P35}. Public and practice-based deliberation plays a crucial role in shaping the emerging norms and practices surrounding AWS, ensuring that they are grounded in real-world applications and public expectations~\cite{P31, P33}. This interaction promotes transparency and accountability in the use of autonomous systems, fostering a better understanding of what constitutes `meaningful' human control~\cite{P31, P33}. Hence, the ethical implications of AWS extend to their operational reliability and safety, as they can enhance military operations by identifying and mitigating software defects, thereby improving overall mission success while adhering to ethical standards~\cite{P46}.

\section{Conclusion\label{section:conclusion}}
The emergence of AWS introduces an important shift in how militaries operate, offering enhanced decision-making capabilities and operational efficiency. AWS present important opportunities, such as improved operational effectiveness, reduced human casualties, and enhanced Human-Machine teaming that can transform military strategies. To fully realise these benefits, while mitigating risks, promoting interdisciplinary collaboration among technologists, ethicists, legal experts, and military strategists is essential. 
Future research should thus focus on the development of robust testing and validation frameworks, improving transparency of the decision-making process, as well as the establishment of comprehensive ethical guidelines for AWS. Additionally, focusing on enhanced system intelligibility, comprehensive training and working towards effective collaboration within the scope of HAT will be very important.

In conclusion, while AWS hold the potential to revolutionise military strategies, their integration into operational contexts must be approached with caution and responsibility. Bridging the existing gap in terms of collaboration between both developers and policymakers is essential in addressing these ongoing challenges.

\bibliographystyle{IEEEtran}
\bibliography{references}

\begin{thebibliography}{10}
\providecommand{\url}[1]{#1}
\csname url@samestyle\endcsname
\providecommand{\newblock}{\relax}
\providecommand{\bibinfo}[2]{#2}
\providecommand{\BIBentrySTDinterwordspacing}{\spaceskip=0pt\relax}
\providecommand{\BIBentryALTinterwordstretchfactor}{4}
\providecommand{\BIBentryALTinterwordspacing}{\spaceskip=\fontdimen2\font plus
\BIBentryALTinterwordstretchfactor\fontdimen3\font minus \fontdimen4\font\relax}
\providecommand{\BIBforeignlanguage}[2]{{%
\expandafter\ifx\csname l@#1\endcsname\relax
\typeout{** WARNING: IEEEtran.bst: No hyphenation pattern has been}%
\typeout{** loaded for the language `#1'. Using the pattern for}%
\typeout{** the default language instead.}%
\else
\language=\csname l@#1\endcsname
\fi
#2}}
\providecommand{\BIBdecl}{\relax}
\BIBdecl

\bibitem{LethalAWS-US}
K.~M. Sayler, ``Defense primer: Us policy on lethal autonomous weapon systems,'' \emph{Congressional Research Service}, 2024.

\bibitem{TAI-XAI}
C.~Vinay, H.~Vikas, S.~Razia, G.~Debshishu, D.~Divyansh, and S.~Biplab, ``A review of trustworthy and explainable artificial intelligence (xai),'' \emph{IEEE Access}, 2023.

\bibitem{TAI-Principles}
B.~Li, P.~Qi, B.~Liu, S.~Di, J.~Liu, .~Pei, J., and B.~Zhou, ``Trustworthy ai: From principles to practices,'' \emph{ACM Computing Surveys}, vol.~55, no.~9, pp. 1--46, 2023.

\bibitem{LethalAWS-UN}
{United Nations General Assembly}, ``Lethal autonomous weapon systems,'' \emph{United Nations General Assembly}, 2023.

\bibitem{AWS_DESIGN}
I.~V. Sancar, ``How can we design autonomous weapon systems?'' \emph{AI and Ethics}, pp. 1--9, 2024.

\bibitem{TRUSTWORTHY_AI}
\BIBentryALTinterwordspacing
B.~Li, P.~Qi, B.~Liu, S.~Di, J.~Liu, J.~Pei, J.~Yi, and B.~Zhou, ``Trustworthy ai: From principles to practices,'' \emph{ACM Comput. Surv.}, vol.~55, no.~9, jan 2023. [Online]. Available: \url{https://doi.org/10.1145/3555803}
\BIBentrySTDinterwordspacing

\bibitem{AWS-Definitions}
T.~Mariarosaria and B.~Alexander, ``A comparative analysis of the definitions of autonomous weapons,'' in \emph{The 2022 yearbook of the digital governance research group}, 2023, pp. 57--79.

\bibitem{AWS-Systemic}
H.~Stephen, ``A cybersystemic view of autonomous weapon systems (aws),'' \emph{Technological Forecasting and Social Change}, vol. 205, p. 123514, 2024.

\bibitem{AWS-Regulation}
{Advisory Council on International Affairs}, ``Autonomous weapon systems: the importance of regulation and investment,'' \emph{Advisory Committee on Public International Law}, 2021.

\bibitem{AWS-Law}
K.~Ahmad and R.~Anandha~Krishna, ``Deployment of autonomous weapon systems in the warfare: Addressing accountability gaps and reformulating international criminal law,'' \emph{Congressional Research Service}, vol.~23, pp. 261--285, 2024.

\bibitem{AWS-UN}
AutoNorms, ``Submission on autonomous weapon systems to the united nations secretary general,'' \emph{SDU Center of War Studies}, 2023.

\bibitem{HAT-MO}
M.~Michael, ``Trusting machine intelligence: artificial intelligence and human-autonomy teaming in military operations,'' \emph{Defense and Security Analysis}, 2023.

\bibitem{HAT-MO-Prop}
M.~Clara, ``Trustworthy human-autonomy teaming for proportionality assessment in military operations,'' in \emph{In 2024 4th International Conference on Applied Artificial Intelligence (ICAPAI) IEEE}, 2023, pp. 1--8.

\bibitem{AWS-ChallengesOpportunities}
S.~Giray and Y.~Cafoglu, ``Challenges and opportunities of autonomous weapon systems (aws) for indigenous defense industries (idi),'' \emph{Security of the Future}, 2024.

\bibitem{AWS-ProsCons}
E.~Amitai and E.~Oren, ``Pros and cons of autonomous weapon systems,'' \emph{Military Review}, 2018.

\bibitem{AWS-Gov}
C.~Esther, K.~Klaudia, and S.~Tim, ``Governing autonomous weapon systems,'' \emph{HCSS Security}, 2020.

\bibitem{AWS-FutureGov}
W.~Joe, ``Developing future capabilities: robotics and autonomous systems,'' \emph{{NATO} Parliamentary Assembly}, 2023.

\bibitem{PRISMA}
M.~J. Page, J.~E. McKenzie, P.~M. Bossuyt, I.~Boutron, T.~C. Hoffmann, C.~D. Mulrow, L.~Shamseer, J.~M. Tetzlaff, E.~A. Akl, S.~E. Brennan \emph{et~al.}, ``Declaraci{\'o}n prisma 2020: una gu{\'\i}a actualizada para la publicaci{\'o}n de revisiones sistem{\'a}ticas,'' \emph{Revista espa{\~n}ola de cardiolog{\'\i}a}, vol.~74, no.~9, pp. 790--799, 2021.

\bibitem{P1}
H.~M. Roff and D.~Danks, ``“trust but verify”: The difficulty of trusting autonomous weapons systems,'' \emph{Journal of Military Ethics}, vol.~17, no.~1, pp. 2--20, 2018.

\bibitem{P2}
A.~Warren and A.~Hillas, ``Friend or frenemy? the role of trust in human-machine teaming and lethal autonomous weapons systems,'' \emph{Small Wars \& Insurgencies}, vol.~31, no.~4, pp. 822--850, 2020.

\bibitem{P3}
M.~Mayer, ``Trusting machine intelligence: artificial intelligence and human-autonomy teaming in military operations,'' \emph{Defense \& Security Analysis}, vol.~39, no.~4, pp. 521--538, 2023.

\bibitem{P4}
B.~Gebru, L.~Zeleke, D.~Blankson, M.~Nabil, S.~Nateghi, A.~Homaifar, and E.~Tunstel, ``A review on human–machine trust evaluation: Human-centric and machine-centric perspectives,'' \emph{IEEE Transactions on Human-Machine Systems}, vol.~52, no.~5, pp. 952--962, Oct 2022.

\bibitem{P5}
J.~Galliott and A.~Wyatt, ``Considering the importance of autonomous weapon system design factors to future military leaders,'' \emph{Australian Journal of International Affairs}, vol.~76, no.~2, pp. 219--244, 2022.

\bibitem{P6}
M.~Firlej and A.~Taeihagh, ``Regulating human control over autonomous systems,'' \emph{Regulation \& Governance}, vol.~15, no.~4, pp. 1071--1091, 2021.

\bibitem{P7}
Y.~Albayram, T.~Jensen, M.~M.~H. Khan, M.~A.~A. Fahim, R.~Buck, and E.~Coman, ``Investigating the effects of (empty) promises on human-automation interaction and trust repair,'' in \emph{Proceedings of the 8th International Conference on Human-Agent Interaction}, 2020, p. 6–14.

\bibitem{P8}
J.~Johnson, ``The ai commander problem: Ethical, political, and psychological dilemmas of human-machine interactions in ai-enabled warfare,'' \emph{Journal of Military Ethics}, vol.~21, no. 3-4, pp. 246--271, 2022.

\bibitem{P9}
M.~C. Horowitz, ``When speed kills: Lethal autonomous weapon systems, deterrence and stability,'' \emph{Journal of Strategic Studies}, vol.~42, no.~6, pp. 764--788, 2019.

\bibitem{P10}
J.~C.~G. Massimiliano Lorenzo~Cappuccio and F.~S. Alnajjar, ``A taste of armageddon: A virtue ethics perspective on autonomous weapons and moral injury,'' \emph{Journal of Military Ethics}, vol.~21, no.~1, pp. 19--38, 2022.

\bibitem{P11}
S.~K. Bhoi, K.~K. Jena, A.~Jena, B.~C. Panda, S.~Singh, and P.~Behera, ``A reputation deterministic framework for true event detection in unmanned aerial vehicle network (uavn),'' in \emph{2019 International Conference on Information Technology (ICIT)}.\hskip 1em plus 0.5em minus 0.4em\relax IEEE, 2019, pp. 257--262.

\bibitem{P13}
R.~Akter, M.~Golam, V.-S. Doan, J.-M. Lee, and D.-S. Kim, ``{IoMT-Net}: Blockchain-integrated unauthorized uav localization using lightweight convolution neural network for internet of military things,'' \emph{IEEE Internet of Things Journal}, vol.~10, no.~8, pp. 6634--6651, April 2023.

\bibitem{P14}
S.~Wasilow and J.~B. Thorpe, ``Artificial intelligence, robotics, ethics, and the military: A canadian perspective,'' \emph{AI Magazine}, vol.~40, no.~1, pp. 37--48, 2019.

\bibitem{P15}
K.~Bogusławski, J.~Nasur, J.~Li, M.~Gil, K.~Wróbel, and F.~Goerlandts, ``A cross-domain scientometric analysis of situational awareness of autonomous vehicles with focus on the maritime domain,'' \emph{IEEE Access}, vol.~10, pp. 50\,047--50\,061, 2022.

\bibitem{P16}
F.~Mancini, S.~Bruvoll, T.~Verhoogt, R.~Wiegers, J.~Melrose, R.~Been, R.~Ernst, K.~Rein, and F.~Leve, ``Securing autonomous and unmanned vehicles for mission assurance,'' in \emph{2019 International Conference on Military Communications and Information Systems (ICMCIS)}, May 2019, pp. 1--8.

\bibitem{P17}
D.~L. Van~Bossuyt, N.~Papakonstantinou, B.~Hale, and R.~Arlitt, ``Trust loss effects analysis method for zero trust assessment,'' in \emph{2023 Annual Reliability and Maintainability Symposium (RAMS)}, Jan 2023, pp. 1--6.

\bibitem{P18}
J.~Haner and D.~Garcia, ``The artificial intelligence arms race: Trends and world leaders in autonomous weapons development,'' \emph{Global Policy}, vol.~10, pp. 331--337, 09 2019.

\bibitem{P19}
S.~Rune and S.~Valaker, ``Mixed-initiative approaches in the design of a trusted shift of coordination forms in air operations: Supporting collaboration to handle loyal wingmen,'' in \emph{2021 International Automatic Control Conference (CACS)}, Nov 2021, pp. 1--6.

\bibitem{P20}
R.~Stensrud, S.~Valaker, and T.~Haugen, ``Interdependence as an element of the design of a federated work process,'' in \emph{2020 IEEE International Conference on Human-Machine Systems (ICHMS)}, Sep. 2020, pp. 1--6.

\bibitem{P21}
P.~Théron and A.~Kott, ``When autonomous intelligent goodware will fight autonomous intelligent malware: A possible future of cyber defense,'' in \emph{MILCOM 2019 - 2019 IEEE Military Communications Conference (MILCOM)}, Nov 2019, pp. 1--7.

\bibitem{P22}
J.~E. Borson and H.~Xu, ``A path dependent approach for characterizing the legal governance of autonomous systems,'' \emph{IEEE Access}, vol.~10, pp. 119\,985--119\,998, 2022.

\bibitem{P23}
N.~Leys, ``Autonomous weapon systems and international crises,'' \emph{Strategic Studies Quarterly}, vol.~12, no.~1, pp. 48--73, 2018.

\bibitem{P24}
M.~S. Ondřej~Rosendorf and M.~Vranka, ``Algorithmic aversion? experimental evidence on the elasticity of public attitudes to “killer robots”,'' \emph{Security Studies}, vol.~0, no.~0, pp. 1--31, 2023.

\bibitem{P25}
A.~Brown, ``Ethics, autonomy, and killer drones: Can machines do right?'' \emph{Comparative Strategy}, vol.~42, no.~6, pp. 731--746, 2023.

\bibitem{P26}
S.~Vallor, ``The future of military virtue: Autonomous systems and the moral deskilling of the military,'' in \emph{2013 5th International Conference on Cyber Conflict (CYCON 2013)}, June 2013, pp. 1--15.

\bibitem{P27}
A.~Roberts and A.~Venables, ``The role of artificial intelligence in kinetic targeting from the perspective of international humanitarian law,'' in \emph{2021 13th International Conference on Cyber Conflict (CyCon)}, May 2021, pp. 43--57.

\bibitem{P28}
J.~Cox and H.~Williams, ``The unavoidable technology: How artificial intelligence can strengthen nuclear stability,'' \emph{The Washington Quarterly}, vol.~44, no.~1, pp. 69--85, 2021.

\bibitem{P29}
R.~J.~M. Boyles, ``Hume's law as another philosophical problem for autonomous weapons systems,'' \emph{Journal of Military Ethics}, vol.~20, no.~2, pp. 113--128, 2021.

\bibitem{P30}
J.~Johnson, ``Deterrence in the age of artificial intelligence \& autonomy: a paradigm shift in nuclear deterrence theory and practice?'' \emph{Defense \& Security Analysis}, vol.~36, no.~4, pp. 422--448, 2020.

\bibitem{P31}
I.~Bode and H.~Huelss, ``Constructing expertise: the front- and back-door regulation of ai’s military applications in the european union,'' \emph{Journal of European Public Policy}, vol.~30, no.~7, pp. 1230--1254, 2023.

\bibitem{P32}
W.~Yang, S.~Wang, X.~Yin, X.~Wang, and J.~Hu, ``A review on security issues and solutions of the internet of drones,'' \emph{IEEE Open Journal of the Computer Society}, vol.~3, pp. 96--110, 2022.

\bibitem{P33}
I.~Bode, ``Practice-based and public-deliberative normativity: retaining human control over the use of force,'' \emph{European Journal of International Relations}, vol.~29, no.~4, p. 990–1016, 2023b.

\bibitem{P34}
M.~Gubrud, ``Stopping killer robots,'' \emph{Bulletin of the Atomic Scientists}, vol.~70, no.~1, pp. 32--42, 2014.

\bibitem{P35}
R.~Brownsword, ``From erewhon to alphago: for the sake of human dignity, should we destroy the machines?'' \emph{Law, Innovation and Technology}, vol.~9, no.~1, pp. 117--153, 2017.

\bibitem{P36}
M.~N. Johnstone and R.~Thompson, ``Security aspects of military sensor-based defence systems,'' in \emph{2013 12th IEEE International Conference on Trust, Security and Privacy in Computing and Communications}, July 2013, pp. 302--309.

\bibitem{P37}
M.~Wagner, ``The dehumanization of international humanitarian law: legal, ethical, and political implications of autonomous weapon systems,'' \emph{Vand. J. Transnat'l L.}, vol.~47, p. 1371, 2014.

\bibitem{P38}
P.~Polishuk and C.~Yin, ``Components for unmanned vehicle systems aircraft/ground/sea/space: Systems, subsystems, components, materials, and other infrastructure equipment and services,'' \emph{Fiber and Integrated Optics}, vol.~32, no. 5-6, pp. 288--323, 2013.

\bibitem{P39}
N.~G. Wood, ``The problem with killer robots,'' \emph{Journal of Military Ethics}, vol.~19, no.~3, pp. 220--240, 2020.

\bibitem{P40}
L.~Johansson, ``Ethical aspects of military maritime and aerial autonomous systems,'' \emph{Journal of Military Ethics}, vol.~17, no. 2-3, pp. 140--155, 2018.

\bibitem{P41}
A.~Brown, ``Ethics, autonomy, and killer drones: Can machines do right?'' \emph{Comparative Strategy}, vol.~42, no.~6, pp. 731--746, 2023.

\bibitem{P42}
M.~Schulzke, ``Autonomous weapons and distributed responsibility,'' \emph{Philosophy \& Technology}, vol.~26, pp. 203--219, 2013.

\bibitem{P43}
E.~D. Reed, ``Truth, lies and new weapons technologies: Prospects for jus in silico?'' \emph{Studies in Christian Ethics}, vol.~35, no.~1, p. 68–86, 2022g.

\bibitem{P44}
M.~D. Żmuda, ``Autonomous weapons of pleasure. media archaeology of automated killing in military and gaming technologies,'' \emph{Culture, Theory and Critique}, vol.~0, no.~0, pp. 1--21, 2023.

\bibitem{P45}
I.~Kerr and K.~Szilagyi, ``Evitable conflicts, inevitable technologies? the science and fiction of robotic warfare and ihl,'' \emph{Law, Culture and the Humanities}, vol.~14, no.~1, p. 45–82, 2018d.

\bibitem{P46}
N.~Gardner, ``Clausewitzian friction and autonomous weapon systems,'' \emph{Comparative Strategy}, vol.~40, no.~1, pp. 86--98, 2021.

\bibitem{P47}
Q.~Zhai and Z.-S. Ye, ``How reliable should military uavs be?'' \emph{IISE Transactions}, vol.~52, no.~11, pp. 1234--1245, 2020.

\bibitem{P48}
V.~L. Foreman, F.~M. Favar{\'o}, J.~H. Saleh, and C.~W. Johnson, ``Software in military aviation and drone mishaps: Analysis and recommendations for the investigation process,'' \emph{Reliability Engineering \& System Safety}, vol. 137, pp. 101--111, 2015.

\end{thebibliography}
\vspace{12pt}

\end{document}